\let\ni=\noindent
\newcommand\ie{{\it i.e.}}
\documentstyle[aps,preprint,amssymb]{revtex}
\begin{document}
\tightenlines

\title{Lagrangian Description of the Variational Equations}
\author{C. M. Arizmendi$^{a}$, J. Delgado$^{b}$, H. N. N\'u\~nez-Y\'epez$^{c}$\footnote{Corresponding author},  A. L. Salas-Brito$^{d}$} 

\address{$^a$ Departamento de F\'{\i}sica, Facultad de
Ingenier\'{\i}a, Universidad Nacional de Mar del
Plata,  Av. J.B. Justo 4302, 7600 Mar del Plata,
Argentina.}
\address{$^{b}$Departamento de Matem\'aticas, 
        Universidad Aut\'onoma Metropolitana-Iztapalapa, 
         Apar\-tado Postal  55-534 Iztapalapa 09340 D.\ F., M\'exico.}
\address{$^{c}$Departamento de F\'{\i}sica, 
        Universidad Aut\'onoma Metropolitana-Iztapalapa, 
         Apar\-tado Postal  55-534 Iztapalapa 09340 D.\ F., M\'exico.}
\address{$^d$Laboratorio de Sistemas Din\'amicos, Universidad
Aut\'onoma Metropolitana-Azcapotzalco, Apartado
Postal 21-267,  Coyoac\'an 04000 D.\ F., M\'exico.}

\maketitle


\begin{abstract}  A variant of the usual Lagrangian scheme is developed which describes both the equations of motion and the variational equations of a  system. The required (prolonged) Lagrangian is defined in an extended configuration space 
 comprising both the original configurations of the system and all the virtual displacements joining any two integral curves. Our main result  establishes that both the Euler-Lagrange equations and the corresponding variational equations of the original  system can be viewed as the Lagrangian vector field associated 
with the first prolongation of the original  LagrangianAfter
discussing  certain features of the  formulation,  we  introduce the so-called inherited constants of the motion and  relate them  to the Noether constants of the extended system.   \end{abstract}


\section {Introduction}

 The variational equations---this is  Whittaker's \cite{Whittaker} terminology---associated with  dynamical systems  are  customarily obtained by linearizing the  equations of motion around a particular solution. Variational  equations   are important for understanding both stability and integrability issues\cite{pla00,steeb,ajp84,uam00,case85}, and interesting since in general relativity and other metric theories of gravitation---as the JBD theory--- they can be regarded as the equations of geodesic deviation, and also because they can be useful for describing linearized gravitation \cite{sussman2,mtw,dicke,robin}. These equations are also useful for studying certain evolution equations\cite{steeb,case85,uam00,case78,matsuno}. Additionally, in chaos theory they are basic for defining  the Liapunov spectra and the related Kolmogorov entropy\cite{pla00,jackson}.  
 
This work  deals   with the variational equations associated to Lagrangian  systems. This is not much of a limitation, however, as
quantum mechanics, geometric control schemes, and field theories can all be described through a Lagrangian function. In all of these the problem of  solving the variational equations is  important \cite{case85,case78,sussman}. We  should also  remember  geodesics in Riemannian manifolds and  their associated Jacobi fields whose properties have inspired much research  and produced  important  results\cite{lanczos,carmo}. 

 Though the  equations of motion of the aforementioned theories stem from a Lagrangian through the Euler-Lagrange equations, the variational equations are normally formulated outside  such scheme,  the fact notwhitstanding that working within the Lagrangian formulation is convenient from a physical and  a mathematical standpoint \cite{pla00,uam00,general,tapia,tapia2}.  It is one of our aims here to  describe a remedy to this situation, that is, we discuss a complete Lagrangian formulation of the variational equations and to take advantage of the description to get constants of motion. 
 A feature of the approach is that the equations can be established making no reference whatsoever to any {\sl specific} solution of the original equations, as is necessarily the case in the standard formulation. This formulation  should also have importance in the theory of the {\it Jacobi fields}  governing the transition from a  geodesic to a nearby one  in the calculus of variations ---\ie\  those vector fields which make the second variation  to vanish identically excepting perhaps for boundary terms.  The notion of Jacobi equations as an outcome of the second variation is in fact fairly more general than this and general formulae for the second variation and generalized Jacobi equations along critical sections have been already considered in the calculus of variations from a sort of  structural, as has been called in \cite{tapia}, point of view (see also \cite{rund66}). 

The paper is organized as follows. In section II we discuss the variational formulation first introduced  in \cite{pla00}. Section III deals with  its the main features of the formulation,  mentioning some of its possible invariances. In section IV we exhibit how to every constant in the original problem it corresponds another one  ---what we call the inherited constant--- valid in the  variational system. In section V we prove Noether theorem for the variational equations, and  establish under what conditions Noether constants can be regarded as inherited ones. We should pinpoint that the  Noether theorem is able to reproduce the known constants associated with symmetries of the original Lagrangian $L$, but that it also implies that to every $n$-parameter symmetry of $L$ there  additionally exist  $n$ new independent conserved quantities in the variational equations---which nevertheless can be trivial in some circumstances. In section VI, we use examples to pinpoint the kind of constants we may obtain by extending symmetries of the original Lagrangian  \cite{rmf02}.
 One employs the simple case of  ignorable coordinates in classical mechanics to explain some further points on the application of Noether's theorem. The other uncovers a conserved vector that is valid in linearized gravitation in a vacuum.    Section VI contains the conclusions and some final remarks.

\section{The Lagrangian for the variational equations}

Let us consider any system  described in terms of a Lagrangian; that is, described using a real-valued function $L$ defined in the tangent bundle $TQ$ of the $N$-dimensional configuration space $Q$ of the system, we additionally assume that $L$ is non degenerate.  
The  equations of motion  of the  system follow from 

\begin{equation} \label{1}
\frac{d}{dt}\left(\frac {\partial L} {\partial \dot q^a}\right)= \frac {\partial L} {\partial q^a}, \quad a=1,2, \dots N.
\end{equation}

\noindent  The solutions of these  Euler-Lagrange equations is the dynamics of the system.  The formulation is described, excepting in section V and in one of the examples of section VI, under the implicit  assumption that the  parameter $t$ is one-dimensional and it is usually referred to as  the time; but it can actually be  any other parameter, as the arc-lenght for geodesics in Riemannian manifolds, or even a finite  set of parameters as  in field theory.   In  such cases the necessary changes are  easily done \cite{carmo,soper,rmf02} as the reader can  verify for herself (see section V and VI for specific examples).

 To describe deviations from the dynamics ---the realm of the Jacobi equations--- let us consider an augmented configuration space  $D$ comprising all the original configurations of the dynamical system plus all the elements of the Jacobi field  that join  two of its  integral curves. The elements of $D$ can hence be coordinatized by pairs $(q^a,\epsilon^a)$ where the $\epsilon^a$ stand for the  ``virtual displacements'' or deviations connecting two  solutions of (\ref{1}). In local coordinates we thus set $q'^a=q^a+\epsilon^a$ and $\dot q'^a=\dot q^a+\dot \epsilon^a$ where $\epsilon^a$ is assumed ``small''.  Formally we may regard $D$ as the double tangent bundle of the configuration manifold of the original system composed with a canonical ``flip'' mapping $\alpha$  reordering the local coordinates as $(q,\epsilon, \dot q, \dot \epsilon)\; \stackrel{\alpha}{\mapsto} \;(q,\dot q, \epsilon, \dot\epsilon)$ \cite{ijtp02}.

 We can now define the new Lagrangian as a function  defined on the tangent bundle $TD$ of the D'Alambert configuration manifold and such that  

\begin{equation} \label{defg}
\gamma(q,\epsilon,\dot q, \dot\epsilon)=\frac {\partial L} {\partial q^a}(q, \dot q) \epsilon^a + \frac {\partial L} {\partial \dot q^a} (q,\dot q)\dot\epsilon^a\equiv {\cal D}_\epsilon L,
\end{equation}

\noindent where we have written the definition in $TD$'s local coordinates, we are using the summation convention and  have taken the opportunity to define the operator ${\cal D}_\epsilon$. Thus $\gamma$ can be regarded as a directional derivative of $L$ along a virtual configuration, or as the effect on $L$ of an operator that ``lifts it a little''  out of the original space\cite{crampin}.  Formally $\gamma$ can be regarded as a {\sl prolongation} \cite{uam00,sussman}   of $L$ \cite{pla00,general}. Note also that $TD$ is the natural domain of definition of $\gamma$ since the virtual displacements are defined in the double tangent bundle of $Q$, and that the mapping $\alpha$ is well defined since the $\epsilon$'s are elements of the tangent space $T_qQ$ at the point $q$ in the configuration space $Q$.    The $N$-dimensional object $\epsilon$ hence plays the role of the variational field associated with the dynamical system's solutions \cite{Whittaker,uam00}. We should pinpoint that despite the fact that  we work in a specific chart of the configuration manifold  all the  results   can be translated into intrinsic language, as has been  done in\cite{uam00,ijtp02}. Be warned that the preliminary  work reported in section 4 of \cite{uam00} has been revised  in \cite{ijtp02}.

Given the Lagrangian $\gamma$, we can immediately write down the $2N$ associated  Euler-Lagrange equations of motion 

\begin{equation} \label{elgep}
\frac {d} {dt} \left( \frac {\partial\gamma}{\partial\dot \epsilon^a}\right)-\frac{\partial\gamma} {\partial\epsilon^a}=0, 
\end{equation}

\ni and

\begin{equation} \label{elgq}
\frac {d} {dt} \left( \frac {\partial\gamma}{\partial\dot q^a}\right)-\frac{\partial\gamma} {\partial q^a}=0. 
\end{equation}

\noindent Using definition (\ref{defg}) in   (\ref{elgep}), we immediately obtain the Lagrange equations of the original system (\ie\ we reproduce Eq.\ (\ref{1}))  and, from (\ref{elgq}), we get their associated linear variational equations \cite{Whittaker,pla00,uam00,case85}

\begin{equation} \label{vareq}
M_{ab} \ddot\epsilon^b + C_{ab} \dot\epsilon^b + K_{ab} \epsilon^b =0, \qquad a=1,2,\dots, N,
\end{equation}

\noindent where the three objects $M$, $C$,  $K$, are defined  by the $N\times N$ matrices

\begin{eqnarray}\label{mat}
 M_{ab}   &=&\left(\frac{\partial^2 L}{ \partial \dot q^a \partial \dot
q^b}\right), \\ 
C_{ab} &=&\left[\frac{d}{ dt} 
   \left( \frac{\partial^2L}{ \partial\dot q^a \partial\dot q^b}
      \right) + \frac{\partial^2 L}{ \partial \dot q^a \partial q^b} -
	\frac{\partial^2 L}{ \partial \dot q^b \partial q^a}\right], \\
K_{ab}&=&\left[\frac{d}{
dt}\left( \frac{\partial^2L}{ \partial\dot q^a \partial q^b} \right)  -
\frac{\partial^2 L}{ \partial 
 q^a \partial q^b}\right], \quad
a,b=1,\dots,N. 
\end{eqnarray}

 The variational equations (\ref{vareq}) and  the  matrices $M$, $C$ and $K$, are {\sl not} explicitly time dependent unless $L$ is it so from the start, a property not found in the standard approaches \cite{Whittaker,case85,case78}. This happens  because we are not  linearizing around a particular solution but taking instead a less local point of view.  Of course, at the  end  is all the same since for solving (\ref{elgep}) we need to insert the solutions of (\ref{elgq}). But  the explicit time-independence of the coefficients in equations of motion (\ref{vareq}) allows  devising  methods for analysing the variational equations analogous to those used in time-independent Lagrangian systems. Such lack of explicit $t$-dependence  can be compared to what happens with the Lagrangian itself which is not  regarded as  time-dependent ---unless it happens to be non autonomous--- despite the fact that  it  depends on $(q^a,\dot q^a)$ and the solutions  are  always explicitly time-dependent.

\section {Some features of the Lagrangian formulation.}

In this section we  pinpoint the main  features of the formulation; for a preliminary or a more mathematical outlook see, respectively, \cite{pla00} and \cite{ijtp02}. 

 As it should be clear by now  the function $\gamma $ is an entirely new Lagrangian which describes the original  system {\sl plus} its response to   perturbations, \ie\ what we have called the ``virtual displacements'' of the system.  It  is thus  useful for studying both matters of stability and, given the relationship of the variational equations with the Painlev\'e test, also of integrability  \cite{Whittaker,pla00,uam00,steeb}. 

 Moreover the Lagrangian formulation  of the variational equations  can be dressed in variational robes by using the function $\gamma$ to define the   action functional 

\begin{equation}
 \Sigma[ q(t), \epsilon(t)]=\int_{t_1}^{t_2}
 \gamma( q,  \dot q, \epsilon,  \dot \epsilon,t)  
dt \label{action}
\end{equation}

\ni of the paths joining two (extended) configurations $(q_1,\epsilon_1)$ and $(q_2,\epsilon_2)$ 
of the system at two given instants of time $t_1$ and $t_2$.  The statement of the variational principle is then just Hamilton's, that is, $\Sigma$ should be extremal when the system follows its actual path in $D$ \cite{pla00,uam00}. In this way we can obtain Eq.\ (\ref{1}) and Eq.\ (\ref{vareq}) as a direct consequence of Hamilton's principle. The variational principle then becomes  a compact  formulation which may allow finding connections between the dynamical properties of the variational fields with other areas of physics or of mathematics \cite{pla00,rund,rau}. 

 As in any Lagrangian formulation, equations (\ref{1}) and (\ref{vareq}) are invariant under arbitrary point transformations ---\ie\ changes of coordinates--- in $Q$,  $\bar{q}^a=f^a(q,t),\; a=1,\dots, N$, where the  $f^a$ are   functions assumed to be  invertible ($\det(df(q))\neq0,\, q\in Q$).   This result can be proved in the standard way \cite{Whittaker,rund}. 

 It is well known \cite{Whittaker,lanczos,LL} that the equations of motion (\ref{1})  remain unchanged if we add to $L$ the total time derivative of any function of $q$ and $t$. This property is reflected in the invariance of Eqs.\ (\ref{elgep}) and (\ref{elgq}), obtained from $\gamma$, under the substitution 

\begin{equation}\label{transf}
\gamma(q,\epsilon,\dot q,\dot\epsilon)\to \gamma(q,\epsilon,\dot q,\dot\epsilon) +\left( \frac {\partial^2 f(q,t)} {\partial q^b\partial q^a} \dot q^b+\frac{\partial^2 f(q,t)} {\partial t \partial q^a}\right) \epsilon^a +\frac {\partial f(q,t)} {\partial q^a} \dot \epsilon^a.
\end{equation}

\ni  Note that the function $f$ is arbitrary except for being required to depend only on $q$ and on $t$. Of course, the equations of motion  also remain unchanged  when the total time derivative of any function, $g$, of $\epsilon$, $q$, and $t$ is added to $\gamma(q,\epsilon,\dot q,\dot\epsilon)$:

\begin{equation}\label{transg}
\gamma(q,\epsilon,\dot q,\dot\epsilon)\to \gamma(q,\epsilon,\dot q,\dot\epsilon) +\frac {d g(q,\epsilon,t)} {dt};
\end{equation}

\ni  the proof of this property follows from the  fact that  the derivative in the right hand side of (\ref{transg}) satisfies identically Eqs.\ (\ref{elgep}) and (\ref{elgq}) \cite{lanczos,LL}.  Any transformation (\ref{transf}) is a particular case of (\ref{transg}) if we choose 

\begin{equation}
g(q,\epsilon,t)=\frac {\partial f(q,t)}{\partial q^a} \epsilon^a.
\end{equation}

\ni This means that the extended system we are describing has a greater invariance than the original one; \ie\ the extended system described by $\gamma(q,\epsilon,\dot q, \dot \epsilon, t)$ is invariant not just under the point transformations in $Q$ explicitly  mentioned above but also under the larger class of point transformations in D'Alambert's configuration space $D$ \cite{pla00,uam00,LL}.

It is to be noted that  the $\epsilon$-derivatives of $\gamma$ reduce to corresponding $q$-derivatives of $L$ as follows

\begin{equation} 
\frac{\partial \gamma} {\partial\dot\epsilon^a}= 
\frac{\partial L} {\partial \dot q^a},\qquad \hbox{and} 
\qquad \frac{\partial  \gamma} {\partial  \epsilon^a}= 
\frac{\partial L} {\partial  q^a}. \label{9}
\end{equation}

\ni As a consequence of Eqs.\ (\ref{9}), $\gamma$ is 
a  first-order homogeneous function of the virtual displacements $\epsilon^a$ and   velocities $\dot \epsilon^a$

\begin{equation}
\gamma= \frac{\partial \gamma} {\partial \epsilon^a}\epsilon^a +
 \frac{\partial \gamma} {\partial \dot\epsilon^a}\dot\epsilon^a, \label{10}
\end{equation}

\noindent and, therefore, it can be also written  in the form

\begin{equation} \label {totalder}
\gamma=\frac {d} {dt} \left( \frac {\partial\gamma} {\partial \dot\epsilon^a} \epsilon^a\right)
\end{equation}
 
\noindent where use have been made of the equations of motion (\ref{elgep}), that is, relation (\ref{totalder}) is valid along the integral curves  associated with Eqs.\ (\ref{1}) and (\ref{vareq}). This equation also implies that the action $\Sigma$ evaluated along such integral curves can be shown to be

\begin{equation} 
\Sigma=  \frac {\partial\gamma} {\partial \dot\epsilon^a} \epsilon^a +\Sigma_0,
\end{equation}

\ni where $\Sigma_0$ is the value of $\Sigma$ at a certain convenient reference time $t_0$.

There are other  relationships that can be written in terms of $\gamma$ and its derivatives. For example, the  Hessian\cite{carmo} of $L$,

\begin{equation} \label{W}
W = \frac {1} {2}  \frac {\partial^2 L} {\partial q^a \partial q^b} \epsilon^a \epsilon^b + \frac {\partial ^2 L} {\partial \dot q^b \partial q^a} \epsilon^a\dot\epsilon^b+ \frac {1} {2}\frac{\partial^2 \gamma}{\partial \dot q^a\partial \dot q^b}\dot \epsilon^a \dot\epsilon^b,
\end{equation}

\ni can be expressed in terms of $\gamma$ as

\begin{eqnarray} \label{hessian}
W &=& \frac {1} {2} {\cal D}_\epsilon \gamma \\
&=&
 \frac{1}{2} \left[ \frac {\delta \gamma} {\delta q^a}\epsilon^a + \frac{d}{dt} \left( \frac {\partial\gamma}{\partial \dot q^a}\epsilon^a\right)\right],
\end{eqnarray}

\noindent where ${\cal D}_\epsilon$ is the operator defined in (\ref{1}), $\delta \gamma/\delta q$ is the functional derivative of $\gamma$ respect $q$ \cite{nash,riesz}. We pinpoint that $W$ can  play an important role in variational problems since its sign  distinguishes between minima, maxima and degenerate critical sections  \cite{carmo,tapia,LL}.  
Note the similitude of the expression for $W$ [Eq.\ (\ref{hessian})] with the definition of $\gamma$, this has interesting consequences for the theory of second variations when the action functionals are defined by first-order Lagrangians \cite{tapia,tapia2}.

\section {Constants of motion in the variational equations.}

Let us consider any constant of motion, $J(q,\dot q)$,  of the Lagrangian vector field Eq.\ (\ref{1}), and let us evaluate it on two of the nearby solutions of Eq.\ (\ref{1}), $q(t)$ and $q'(t)=q(t)+\epsilon(t)$. The difference between such constant quantities evaluated in nearby trajectories, $j(q, \dot q, {\epsilon}, \dot{\epsilon})\equiv 
J( {q}',\dot{ q}')-J({ q},\dot{ q})$,  
is also trivially a constant, 
 
\begin{equation} 
\frac{d j({ q,  \epsilon}, \dot q, \dot{\epsilon})} {dt}=\frac{d} {dt}
\left[ J({ q}',\dot {q}')-J({ q},\dot{ q}) \right] =0. \label{17} 
\end{equation}

\noindent  This  constant, $j[q,{ \epsilon}]$, which we call  an {\sl inherited constant}\cite{uam00}, can be  also expressed  as 
 
\begin{equation} \label {incons}
j(q,{ \epsilon},\dot q, \dot \epsilon) = {\cal D}_\epsilon J= \left(\frac{\partial J} {\partial q^a}\epsilon^a+ \frac{\partial 
J} {\partial \dot q^a}\dot\epsilon^a \right). \label{18}
\end{equation}

\ni \ie\ as a sort of directional derivative of $J$ along a virtual  configuration. Equation (\ref{18}) tells us how, given both a solution of equations (\ref{1}) and any one of its constants of motion, we can obtain a constant of motion for equations (\ref{vareq}).   This result  establishes  a direct relationship between integrals of motion in the variational equations, like $j$, to constants in the original equations (\ref{1}).  Related  results  are discussed in \cite{case85}. We need to pinpoint that, as in some non-linear evolution equations the integrals of motion $J[q(t)]$ are functionals \cite{nash,riesz} ---and not functions--- of the solutions of  (\ref{1}), then  the constant  $j$   has to be regarded also  as a functional of the Jacobi fields $\epsilon(t)$, given  by \cite{uam00,case78}

\begin{equation}
j[\epsilon(t)]=\int \frac{\delta J[q(t)]}{ \delta q(\tau)} \epsilon(\tau)\; d\tau, 
\label{19}   
\end{equation}

\noindent as has been shown in \cite{case85}; in Eq.\ (\ref{19}) ${\delta J[q(t)]/ \delta q(\tau)}$ stands for the functional derivative of  the  functional $J[q(t)]$.

At this point we  recall the  relationship between  symmetries of $L$ and the existence of integrals of motion in Lagrangian systems. This relation is summarized in Noether's theorem, which we, in the next section, proceed to extend   to the existence of constants of motion in the variational equations. Such extension casts (\ref{incons}) in a different and perhaps more interesting form by relating it to a symmetry of the original Lagrangian $L$. But, as we mentioned before, the symmetries of the variational equations are larger than the symmetries of the original Lagrangian. Thus the Noether's theorem  relates any continuous symmetry of $\gamma$---not necessarily coming directly from one of $L$---to a constant of motion in the variational equations.

\section{Symmetries and Integrals of Motion.}

In the previous section we established the existence of what we called inherited constants of motion. As every isolating constant in a Lagrangian system come from a symmetry transformation \cite{LL,rhor}, this has exhibited that to every symmetry of $L$ there exist a related integral of motion in the variational equations.  This  also shows that any symmetry of $L$ can also be regarded as a symmetry of $\gamma$;  the converse of this statement is not necessarily true.

In this section we  formalize the just mentioned relation between constants of motion and symmetries of $\gamma$, making explicit the relation between the inherited constants of motion and the generators of the symmetry transformations of $L$. These results can be of particular importance in field-theoretic or in non-linear evolution equation applications \cite{case85,case78,tapia2,matsuno}. Thinking on  such field theoretic applications, in this section we assume that $L$ depends on $n$ parameters $t^\mu$ and not on a single one as we do in the rest of the paper. The $n$-parameters are further assumed to belong to a given $n$-dimensional differentiable manifold $V_n$ as it would happen if they were   points in the space-time continuum.

Thus,  let us suppose that the Lagrangian $\gamma$ is invariant under the $r$-parameter continuous group of transformations

\begin{equation}\label{tg}
 \bar{q}^a= F_{\bar{q}}^a(t,q,\epsilon, w),\quad\bar{\epsilon}^a=F^a_{\bar{\epsilon}}(t,q,\epsilon,w),\quad \bar{t}^\mu=F^\mu_{\bar{t}}(t,q,\epsilon,w),
\end{equation}

\ni in which the $\omega^s,\; (s=1,\dots, r)$ denote the $r$ parameters of the group  furthermore assumed independent of each other.  The functions $F_{\bar{q}}^a$ and $F_{\bar{t}} $ are parameterized by $\omega^s\in I$ (where $I$ is the set of parameters of the transformation group),   and such that $\omega^s=0,\; s=1\dots r,$ corresponds to the identity transformation, \ie\ the group is continuously connected with the identity.  The transformations (\ref{tg}) can  be also written in  the infinitesimal form

\begin{eqnarray}\label{itg}
  \bar{q}^a=q^a + \zeta^a_{s}(t,q,\epsilon)\omega^s,\\
\bar{\epsilon}^a=\epsilon^a + \eta^a_{s}(t,q,\epsilon)\omega^s,\\
\bar{t}^\mu=t^\mu+ \xi^\mu_s(t,q,\epsilon)\omega^s,
\end{eqnarray}

\ni where $\xi_s(t,q,\epsilon)$, $\zeta^a_{s}(t,q,\epsilon)$ and  $\eta^a_{s}(t,q,\epsilon)$ are the  infinitesimal generators  of the symmetry transformation, given by

\begin{eqnarray}
\zeta^a_{s}&=& \left(\frac{\partial F_{\bar{q}}^a}{\partial \omega^s}\right)_{\omega^s=0},\label{igtg1}\\
\eta^a_{s}&=& \left(\frac{\partial F^a_{\bar{\epsilon}}}{\partial \omega^s}\right)_{\omega^s=0},\label{igtg2}
\\
\xi^\mu_s&=& \left(\frac {\partial F^\mu_{\bar{t}}} {\partial \omega^s}\right)_{\omega^s=0}\label{igtg3}.
\end{eqnarray}

  That under the infinitesimal transformations (\ref{igtg1}--\ref{igtg3}) the system described by $\gamma$ remain invariant means that the  action $\Sigma$ [Eq.\ (\ref{action})] must behave under the   symmetry transformations as \cite{rund}

\begin{equation}\label{condition}
\gamma \left(
\bar{q}^a,
\frac{\partial \bar{q}^a} {\partial \bar{t}^\mu}, 
\bar{\epsilon}^a, 
\frac{\partial \bar{\epsilon}^a} {\partial \bar{t}^\mu},
\bar{t}^\mu
\right)
\det \left(\frac{\partial \bar{t}^\mu}{\partial t^\nu} \right) = \gamma\left( 
{q}^a,
{\epsilon}^a,
\frac{\partial {q}^a} {\partial {t}^\mu},
\frac{\partial {\epsilon}^a} {\partial {t}^\mu}, 
{t}^\mu 
\right).
\end{equation}  

The proof of the existence of the conserved quantities ---details missing in the following proof can be filled in following the discussions in \cite{rund,rhor}--- is  direct.  Begin with (\ref {condition}),  differentiate it respect to  $\omega^s$ noting that the right hand side is  independent of such parameters, substitute $\omega^s=0$ after differentiation, and  employ the equations of motion, to obtain 

\begin{equation}
\frac{\partial j^\mu_a} {\partial t^\mu}=0,
\end{equation}

\ni where the divergenceless tensor $j^\mu_a$ is given by

\begin{equation}\label{j}
j^\mu_a= \gamma\xi^\mu_a+\frac{\partial\gamma} {\partial q^s_{,\mu}}\left(\zeta^s_a -q^s_{,\nu} \xi^\nu_a\right) + \frac{\partial\gamma} {\partial \epsilon^s_{,\mu}}\left(\eta^s_a -\epsilon^s_{,\nu} \xi^\nu_a\right),
\end{equation}

\ni where we use the usual notation $\epsilon_{,\nu}\equiv \partial {\epsilon}/\partial t^\nu$. The result (\ref{j}) is  the main consequence of the Noether's theorem, establishing the connection between symmetries and conserved quantities at the level of the variational equations of a Lagrangian dynamical system. Let us point out that Eq.\ (\ref{j}) was to be expected since $\gamma$ is indeed a Lagrangian.  
 
 As the  number of symmetries of the variational equations is larger than that of the original equations, we may ask under what conditions  the Noether's constants  (\ref{j}) reduce to the case of an inherited constant (\ref{incons}). If we regard the original constant of motion $J^\mu_a$ as  coming also from a Noether symmetry, then

\begin{equation}\label{JN}
J^\mu_a= L\xi^\mu_a + \frac {\partial L} {\partial q^s_{,\mu}} \left(\zeta^s_a-q^s_{,\nu} \xi^\nu_a\right)
\end{equation}

\ni the variational equations constant $j^\mu_a$  can always be written in terms of $J^\mu_a$ and of the operator ${\cal D}_\epsilon$ as

\begin{equation}\label{jinh}
j^\mu_a={\cal D}_\epsilon J^\mu_a + \frac {\partial \gamma} {\partial \epsilon^k_{,\mu}} \left( \eta^k_a - \epsilon^k_{,\beta} \xi^\beta_a \right)  - L \;{\cal D}_\epsilon \xi^\mu_a - \frac {\partial \gamma} {\partial \epsilon^k_{,\mu}}\; {\cal D}_\epsilon \left( \zeta^k_a - q^k_{,\beta} \xi^\beta_a \right). 
\end{equation}

\ni where we have assumed the identity of the $\zeta^s_a$ and of the $\xi^s_a$  appearing in Eq.\ (\ref{JN}) with those in Eq.\ (\ref{jinh}). It is quite clear then that not every constant $j^\mu_a$  can be regarded as an inherited constant. For this relation to be true two things are needed: i) that $\eta^b_a=0$ \ie\ that the transformation does not involve directly the $\epsilon^a$,  and ii) that one or the other (or both) of the generators, $\zeta^b_a$ and $\xi^b_a$, be constants.
 
\section {Two Examples.}

 In this section we give two examples of the use of  Noether's theorem in the variational equations.   

\subsection {Ignorable coordinates.}  

Let us first  consider  a $N$-degrees of freedom mechanical system described by the  Lagrangian, 

\begin{equation}\label{lag}
L(q,\dot q)=m_{ab}\dot q_a \dot q_b-V(q).
\end{equation}

 \ni We assume this Lagrangian does not depend on the specific coordinate $q^A$. The function $\gamma$ is then

\begin{equation}\label{gam}
\gamma= \left(\frac {1} {2}\frac {\partial m_{ab}} {\partial q_s} \dot q_a \dot q_b - \frac {\partial V} {\partial q_s} \right)\epsilon_s + m_{ab} \dot q_a \dot \epsilon_b.
\end{equation} 

\ni and it neither depends on $q^A$. A one-parameter group of symmetries of both (\ref{lag}) and (\ref{gam}) is  the translation

\begin{equation}\label{trans}
 \bar{q}^a=q^a,\;\, a\neq A,\quad\hbox{ and }\quad   \bar{q}^A=q^A +w,
\end{equation}

\ni which is applied  on $Q$, the original configuration space.  The conserved quantity coming from such invariance  is $q^A$'s conjugate momentum $p_A={\partial L} /{\partial \dot q^A} $. 
We are now going to illustrate the relation between $p_A$ above [see  Eq.\ (\ref{JN})] and  the constant in the variational equations [Eq.\ (\ref{jinh})].  The precise answer  depends on how  the transformation  (\ref{trans}) is applied on $D$. As it is  easy to convince oneself, the following two are the only independent possibilities allowed.
 
\begin {itemize}
\item[1.] First, we apply (\ref{trans}) to $q^s $ and not to the ``virtual displacements'' or the time, \ie\ $\bar{\epsilon}^a=\epsilon^a$, $\bar{t}=t$, ($\zeta^A=1$, $\zeta^a=0,\, a\neq A$; $\eta^a=0$; $\xi^a=0$). In this instance the conserved quantity is
\begin{equation}
j^{(1)}_A= \frac {\partial \gamma} {\partial \dot q^A}=\frac {\partial m_{Ab}}{\partial q^a}\, \dot q^b \epsilon^a + m_{Ab}\dot\epsilon^b, 
\end{equation}

\ni and can be written as the inherited constant associated with $p_A$, \ie\  $j^{(1)}_A={\cal D}_\epsilon p_A$.

\item[2.]  In the second case, transformation (\ref{trans}) is  just applied to  the $\epsilon$'s and not to the $q's$ or the $t$. We have $\bar{q}^a=q^a$; $\bar{\epsilon}^A=\epsilon^A +w$, $ \bar {\epsilon}^a=\epsilon^a,\, a\neq A$;  $\bar{t}=t$ (\ie\ $\eta^A=1$, $\eta^a=0,\, a\neq A$; $\zeta^a=\xi^a=0$).   The conserved quantity is  just the momentum conjugate to $q^A$, $j^{(2)}_A =  {\partial \gamma}/  {\partial \dot \epsilon^A} = m_{Ab}\dot q^b = p_A$. 

\end{itemize}

  It is worth emphasizing that this is the typical behaviour when we apply  a symmetry of $L$  to $\gamma$.  We always obtain a new  conserved quantity  $j^{(1)}$, and a pre-existing (in  $L$) one  $j^{(2)}$. 
We pinpoint that it is not necessary that the constant $j^{(1)}$ be   inherited, as it  happened in the present example;  this might not even be  possible.  We must keep in mind  that besides the inherited symmetries, $\gamma$ may have other symmetries of its own and therefore other associated constants. 

\subsection {Conserved quantities in linearized gravitation in a vacuum.}  

  Let us investigate conserved quantities in a first-order theory of gravitation in a vacuum. The Lagrangian density \cite{robin}

\begin{equation}\label{gr}
L=\frac {1} {2} R^{a\mu b\nu} x_{a,\mu} x_{b,\nu},
\end{equation}

\ni with $R^{a\mu b\nu}$ the Riemann tensor, describes vacuum general relativity. The associated density  is

\begin{equation}\label{lgr}
\gamma=\frac {1} {2} \frac {\partial R^{a\mu b\nu}} {\partial x_c} \epsilon_c x_{a,\mu} x_{b,\nu} + R^{a\mu b\nu} \epsilon_{a,\mu} x_{b,\nu}.
\end{equation}

\ni this Lagrangian describes linearized gravitation in a vacuum.

It can be easily realized that  under the following one parameter group of transformations of the spacetime

\begin{equation} \label{camp}
\bar{x}^a=x^a + x^a w, 
\end{equation}

\ni the Lagrangian (\ref{gr})  changes as $\bar{L}=  L/({1+w})^2$. So, though it is not strictly invariant, $L$  ends multiplied by a constant,   thus there is a Noether  tensor  $\theta^\mu=- R^{a \mu b \nu} x_{a, \mu} x_{b, \nu} x^\mu/2 + R^{a \mu b \nu} x_{a,\mu} x^b$ which is not very interesting since it vanishes identically. Furthermore, the density $\gamma$ admits the following two (quasi) symmetry transformations:

\begin{itemize}
\item [1.] For the first one, we just apply transformation (\ref{camp}), not changing the $\epsilon$'s in any way. The Lagrangian $\gamma$ transforms as $\bar {\gamma} = (1+w)^{-3}\, \gamma$, as a result the quantity

\begin{eqnarray}\label{lg1}
&j^\mu &= \gamma x^\mu + \frac {\partial\gamma} {\partial x^a_{,\mu}} (x^a - x^a_{,\beta} x^\beta) - \frac {\partial\gamma} {\partial \epsilon^a_{,\mu}} ( \epsilon^a_{,\beta} x^\beta) \nonumber \\
&=& {R_a}^{\mu b \nu} \epsilon_{b, \nu} x^a
\end{eqnarray}

\ni is divergenceless\cite{robin}.

\item [2.] The second invariance does not affect the spacetime 
coordinates $x^a$ just transforms  the ``virtual'' displacements $\epsilon$ 
as

\begin{equation}
\bar{\epsilon}^a= \epsilon^a + \epsilon^a w,
\end{equation}

\ni under this change the Lagrangian changes as $\bar{\gamma}= (1+w)^{-1}\, \gamma$ and as a result the quantity $j^\mu= {\partial \gamma}/ {\partial \epsilon^a_{,\mu}} \epsilon^a
=0$ is also ---trivially--- divergenceless. This invariance does not give any useful new information.
\end{itemize}

 The first divergenceless tensor (\ref{lg1}), which do not stem from any constant in the original spacetime, is valid in linearized gravitation and most important remain constant irrespective of what spacetime we are linearizing around. 

\section{Conclusions}

The  Lagrangian description  discussed in this paper has led to  the uncovering of various properties of the  variational equations of Lagrangian systems, and  allowed  the application  of Noether's theorem to them, thus relating  the symmetries of a Lagrangian  not only to its constants of motion but also to constants in its variational equations. These conclusions  stem from the introduction of the Lagrangian $\gamma$ which, from a mathematical point of view, is the prolongation of $L$ \cite{uam00,ijtp02}. It is to be emphaziced that our results are valid for field theories and non-linear evolution equations and can be also extended to the higher order Lagrangians which found application in relativistic theories\cite{pla00,tapia}. Moreover, we can apply the techniques presented here in the   framework of first-order variational principles on fibered manifolds and their jet prolongations \cite{tapia2,sarda}. Our formulation may also possibly be used to study the so-called geoodular structure (a mathematical scheme that defines  curvature in terms of properties of congruences of geodesics and their variational equations   in affinely connected manifolds\cite{neste}).  The extension of Noether's theorem to the variational equations presented here may also have some bearings in the study of  the  evolution of Stokes waves\cite{matsuno}, and can be also used in  geometric control theory \cite{uam00,sussman}.

On the other hand, the relevance of the variational equations for determining both stability and integrability properties,  makes this formulation important for  the investigation   of periodic orbits \cite{steeb} and  of solitonic solutions of nonlinear equations \cite{matsuno}. Our Lagrangian description of the Jacobi equations  can  be durthermore casted in  Hamiltonian form.   Moreover, as the example B of section VI shows, the formulation lends itself to describe linearized gravitation. 
Let us mention that the mathematical setting of the  concepts behind our  formulation are    explained more formally in \cite{ijtp02} for the case of discrete Lagrangian mechanics.
 

\bigskip 
\acknowledgements { 

This work was partially supported by research grants  from the Office of Naval Research, the  Universidad Nacional Aut\'onoma de M\'exico (PAPIIT-IN) and from  CONACyT (32167-E).    We  thank  D. C. Robinson (King's College, London) and  R. Sussman (ICN-UNAM)  for their helpful remarks or suggestions. A.L.S.-B. and H.N.N.-Y. want to thank their  friends B.\ Ch.\ Caro, R.\ Micifuz, F.\ C.\ Cucho,  U.\ Kim, C. Ch. Ujaya, P.\ Weiss,  E.\ Hera, J.\ E.\ Juno, R.\ Zeus, K. Bielii,  P.M. Schwarz, C. F. Quimo, P. Sabi, and M.\ Botitas for their encouragement. Last but not least we dedicate this work to the memories of M.\ Osita (1990--2001), Ch. Shat (1991--2001), and F. C. Bonito (1987--2002). 
}

\end{document}